\documentclass[%
 reprint, aps,
 amsmath,amssymb,
prb,
]{revtex4-1}

\usepackage{graphicx}
\usepackage{dcolumn}
\usepackage{bm}
\usepackage{hyperref}
\usepackage[colorinlistoftodos]{todonotes}
\usepackage{ulem}
\graphicspath{{Figures/}}
\begin{document}

\preprint{APS/123-QED}

\title{Optical forces in nanorod metamaterial}

\author{Andrey A. Bogdanov$^{1,2}$}
\email{bogdanov@ioffe.mail.ru}
\author{Alexander S. Shalin$^{1}$}
\author{Pavel Ginzburg$^{3}$}

\affiliation{
$^{1}$    ITMO University, 197101 St.~Petersburg, Russian Federation \\
$^{2}$    Ioffe Institute, 194021 St.~Petersburg, Russian Federation \\
$^{3}$    School of Electrical Engineering, Tel Aviv University, 69978 Tel Aviv, Israel}

\date{\today}

\begin{abstract}
 Optomechanical manipulation of micro and nano-scale objects with laser beams finds use in a large span of multidisciplinary applications. Auxiliary nanostructuring could substantially improve performances of classical optical tweezers by means of spatial localization of objects and intensity required for trapping. Here we investigate a three-dimensional nanorod metamaterial platform, serving as an auxiliary tool for the optical manipulation, able to support and control near-field interactions and generate both steep and flat optical potential profiles. It was shown that the 'topological transition' from the elliptic to hyperbolic dispersion regime of the metamaterial, usually having a significant impact on various light-matter interaction processes, does not strongly affect the distribution of optical forces in the metamaterial. This effect is explained by the predominant near-fields contributions of the nanostructure to optomechanical interactions. Semi-analytical model, approximating the finite size nanoparticle by a point dipole and neglecting the mutual re-scattering between the particle and nanorod array, was found to be in a good agreement with full-wave numerical simulation. In-plane (perpendicular to the rods) trapping regime, saddle equilibrium points and optical puling forces (directed to the excitation light source along the rods), acting on a particle situated inside or at the nearby the metamaterial, were found. The auxiliary metamaterials, employed for optical manipulation, provide additional degrees of freedom in flexible nano-mechanical control and could be employed in various cross-disciplinary applications. 
\end{abstract}

\maketitle


\section{Introduction \label{sec:intro}}
The ability to control mechanical motion of micro- and nano-scale particles with focused laser beams is an essential tool, being a paramount for a wide range of applications, related to bio-physics, micro-fluidics, optomechanical devices and more.\cite{juan2011plasmon,Dienerowitz2008,marago2013optical,grier2003revolution} Being first proposed and demonstrated by A. Ashkin,\cite{ashkin1970acceleration} the classical optical tweezers are nowadays a rapidly developing area of fundamental and applied research.

One of the promising and already conceptually proven approaches for improving performances of the optical manipulation schemes is to employ various auxiliary nanostructures, especially plasmonic ones. \cite{juan2011plasmon} The key idea of the plasmonic tweezers is to utilize strong light-matter interaction between nanostructured metals and focused laser beams. Noble metals, having a negative permittivity in the optical and infrared spectral ranges, support localized plasmon resonances enabling enhancement and control of near-fields at their vicinity.\cite{novotny2012principles,berkovitch2012nano} In particular, the creation of strong intensity gradients is beneficial for obtaining substantial optical forces, which is important, for example, for achieving molecular manipulation.\cite{shoji2014plasmonic} 

Plasmonic nanostructures with subwavelength light concentration could be employed for obtaining new optomechanical effects, i.e. accelerating nanoparticles in an arbitrary direction (in relation to the light propagation direction),\cite{shalin2013plasmonic} or for creating nano-modulators of plasmonic signals.\cite{shalin2014nano} 

Arrays of antennas and their integrations in photonic circuitry,\cite{juan2011plasmon,lin2014trapping} employed as auxiliary tools for optical trapping, were shown to outperform classical schemes (focused lasers in homogeneous media, e.g. liquid solutions) both in terms of spatial localization and optical power required per trapped particle.Antenna arrays were further extended for multifunctional platforms, enabling trapping, stacking, and sorting. \cite{roxworthy2012application} However, isolated plasmonic structures create limited number of hot spots (local enhancement of intensity) and are usually restricted to two-dimensional geometries. These constrains set significant limitations on the flexibility of optical manipulation by reducing potential degrees of freedom, available for optomechanical control. On the other hand, three-dimensional artificially created nanostructures or metamaterials (see, e.g., Ref.~\onlinecite{cai2010optical}) could provide additional benefits and flexibility by configuring near-field interaction in large volumes.

Hyperbolic metamaterials \cite{shekhar2014hyperbolic} are one class of artificially created electromagnetic structures, capable to enhance efficiencies of various light-matter interaction processes owning to the unusual dispersion of eigenmodes, supported by the structure -- namely its hyperbolic dispersion. Among various possible designs of this type of metamaterials it is worth mentioning composites made of vertically aligned nanorods,\cite{atkinson2006anisotropic,ginzburg2013manipulating} periodic metal dielectric layers,\cite{krishnamoorthy2012topological} and semiconductor quantum structures.\cite{hoffman2007negative,ginzburg2008nonmetallic}	While the far-field interactions of waves with hyperbolic composites were proven to be well characterized in terms of the effective medium approximation, this description could be questioned if near-field mediated processes are involved  [see, e.g., Ref.~\onlinecite{ginzburg2013self}].	

The general criterion for estimating impact of near-field contributions to an interaction is based on comparison of k-vector spectra with reciprocal vector of the metamaterial lattice. For example, scattering from objects within hyperbolic metamaterials involves consideration of the near-field effects  [see, e.g., Ref.~\onlinecite{shalin2015scattering}]. 

Analysis of near- and far-field contributions to optical force, acting on objects embedded in the nanorod metamaterial is the central topic of the manuscript. In particular, optical forces, acting on nano-sized spherical particle embedded inside the metamaterial assembly, are investigated both numerically and by using a semi-analytical approach, considering the finite size nanoparticle as a point dipole and neglecting re-scattering between the particle and nanorod array. The impact of the finite structure of the metamaterial unit cell and the relative arrangement of the particle in respect to it was analyzed as a function of the system's geometry and frequency of incident illumination. A semi-analytical approach based on dipole near-field interaction is developed and shown to be in a good agreement with results of the full-wave numerical analysis. The interplay between near- and far-field effects in the context of effective medium approximation is discussed. 

The manuscript is organized as follows. In Sec.~\ref{sec:effective_medium} the nanorod metamaterial platform is introduced and its optical properties are analyzed. In	Sec.~\ref{sec:results} the distribution of optical forces is investigated numerically. A semi-analytical model is introduced and compared with the full-wave numerical simulation. Finally, Sec.~\ref{sec:conclusion} summarizes the major results.

  \begin{figure}[t]
   \centering
  \includegraphics[width=0.95\linewidth]{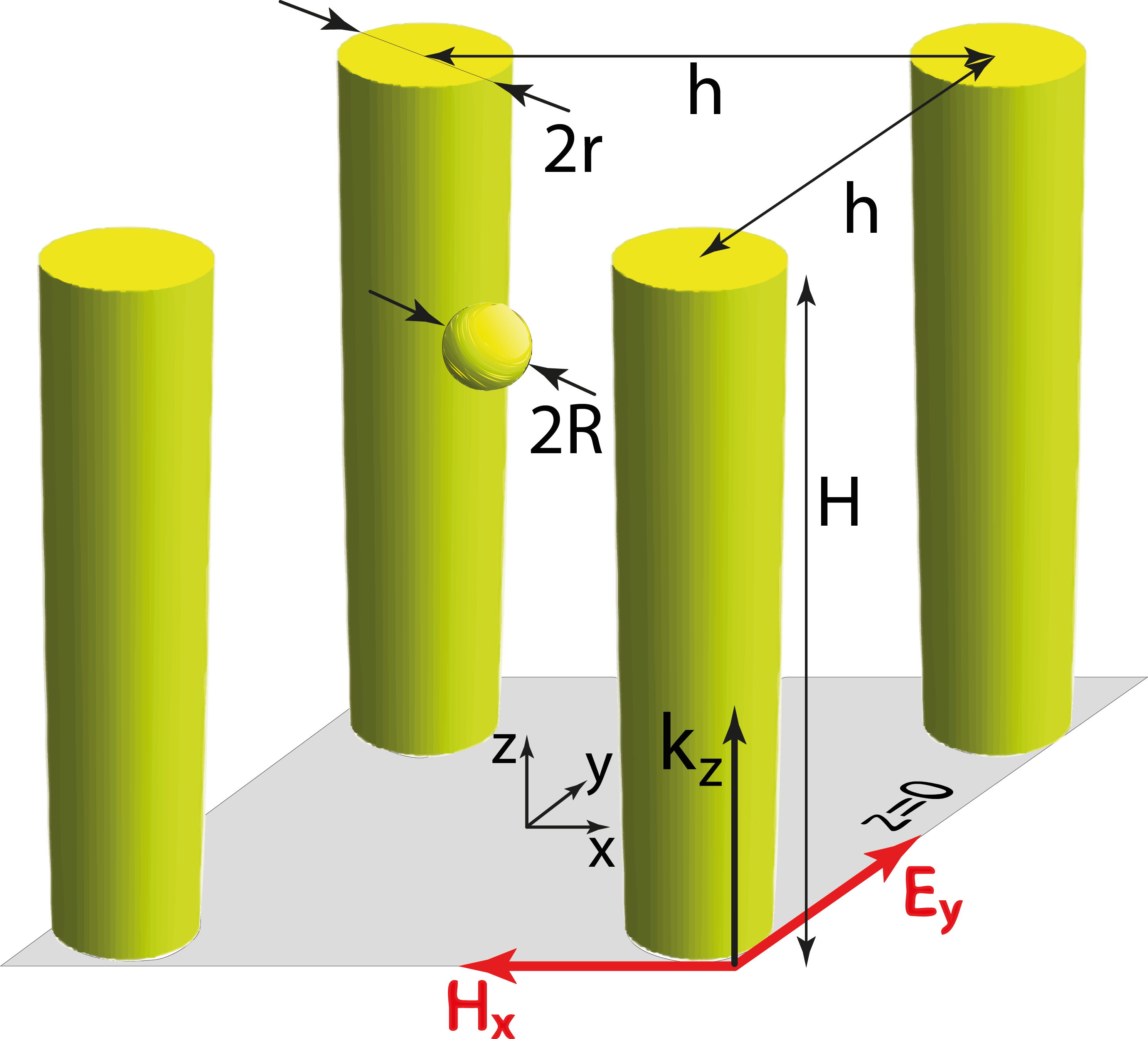} 
  \caption{(color online) Schematics of the nanorod metamaterial with a spherical nanoparticle inside. Radius of the nanoparticle is $R$~=~5~nm. Height and radius of the nanorods are $H$~=~350~nm and $r$~ = ~15~nm, respectively, the period is $h$~=~60~nm.}
\label{design}
\end{figure}

\section{Nanorod Metamaterial: \  \ far-field characteristics \label{sec:effective_medium}}

The geometry under investigation is schematically represented in Fig.~\ref{design} -- it shows an array of vertically aligned gold nanorods and a gold nanoparticle placed inside it. Material parameters of the constitutive elements were taken from widely used sources.\cite{Johnson1972} The parameters of the structure are indicated in the figure caption. While the nanorods in this model are situated in vacuum, substrate effects and host material filling the space between the rods could be taken into account straightforwardly. Similar structures have already found use in various multidisciplinary applications, among them bio-sensing,\cite{kabashin2009plasmonic} enhancement of nonlinearities,\cite{wurtz2011designed} acoustic waves detection,\cite{yakovlev2013ultrasensitive} thin optical elements\cite{ginzburg2013manipulating} and others. The key properties of this auxiliary nanostructure leading to enhanced performance are large surface area and unusual collective optical response of the system, enabling control over both far- and near-field interactions. Hence, investigation of optical forces, mediated by nanorod metamaterials, has a profound potential interest.


      \begin{figure}[t]
   \centering
   \includegraphics[width=0.95\linewidth]{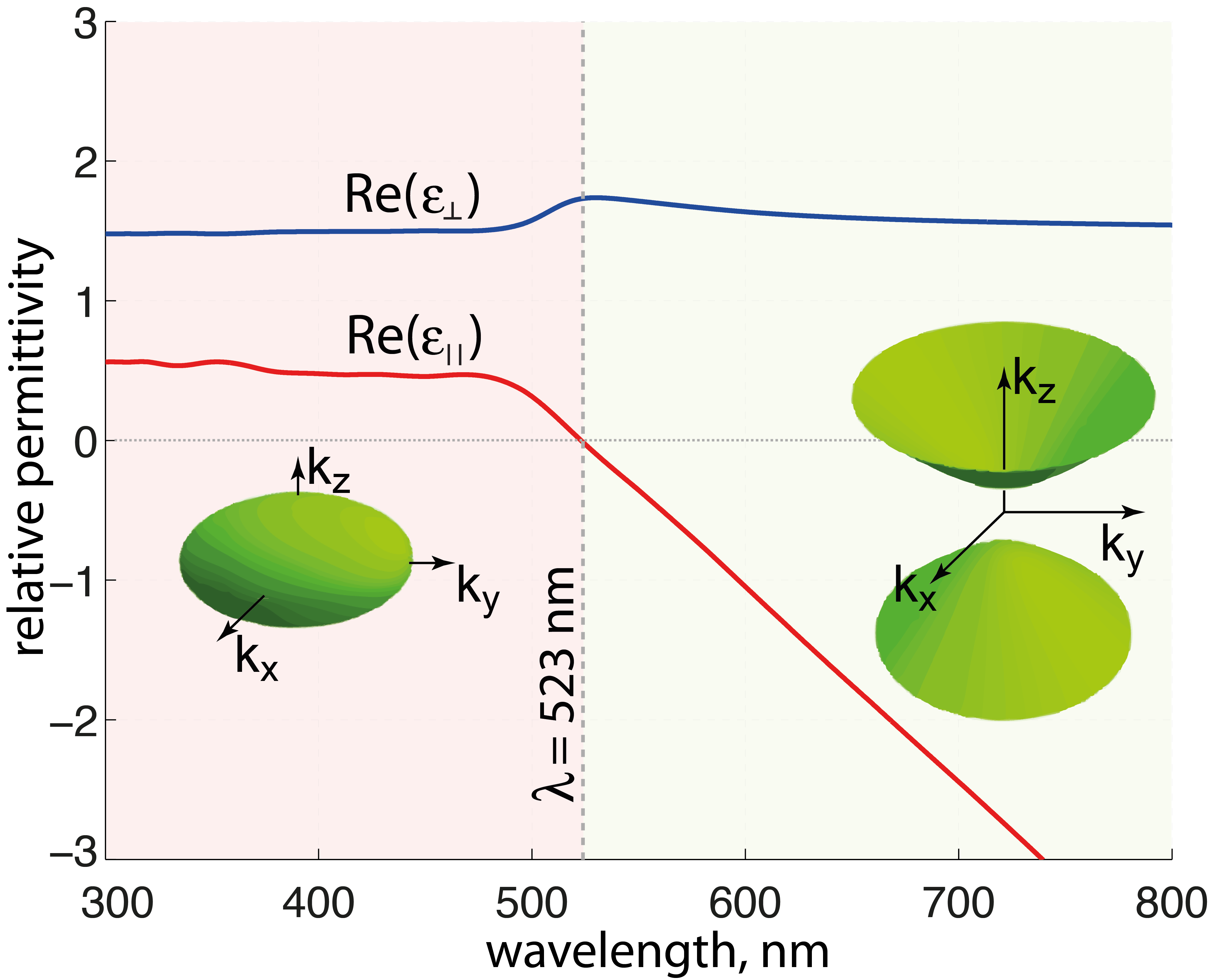} 
   \caption{ (color online) Frequency dependence of the effective tensor components (real parts) $\varepsilon_{\bot}$ and $\varepsilon_{\|}$ of  the nanorod metamaterial. Dashed line shows the transition between elliptic and hyperbolic dispersion regimes of the metamaterial -- epsilon near zero (ENZ) point. Insets show characteristic  shapes of iso-frequency surfaces, corresponding to the dispersion regimes.}
   \label{fig_eff}
\end{figure}

Far-field interactions between electromagnetic waves and metamaterials, under certain circumstances, can be described within the effective medium approximation. The main idea of this homogenization procedure is to average the electromagnetic field over a unit cell of the structure. Therefore, the field inside the structure is assumed to be uniform. In the context of optical forces, as it will appear in the next section, the non-uniformities play a major role and, in fact, predefine the spatial structure of optical potentials. Recently, a phenomenological approach taking into account the finite size of the metamaterial unit cell was proposed.\cite{Shalin2015pulling} Inclusion of a depolarization volume around optically manipulated particles enabled investigations of far-field contributions to optical forces. However, near-field interactions, being strongly dependent on a specific metamaterial design, were not included explicitly.

One of the key properties of hyperbolic metamaterials, making them attractive for electromagnetic applications, is their unique ability to support an unusual regime of dispersion caused by having permittivity tensor components of opposite signs  ($\varepsilon_\bot\varepsilon_\|<0$). An immediate implication of this hyperbolic dispersion regime is the high density of photonic states, available for both emission and scattering.\cite{jacob2010engineering,iorsh2015compton}

     \begin{figure*}[htbp]
   \centering
   \includegraphics[width=0.9\linewidth]{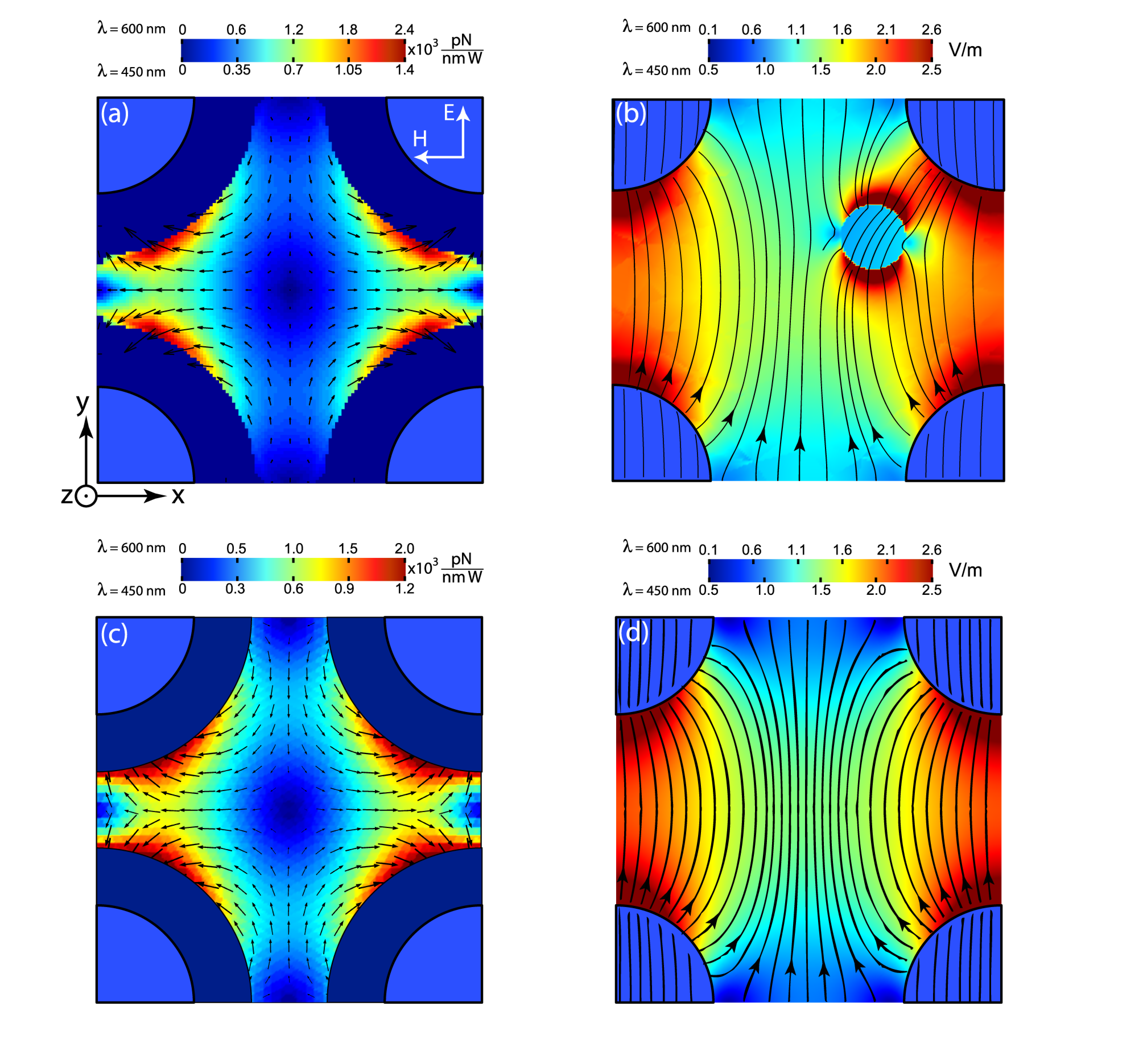} 
   \caption{ (color online)  (a) Lateral optical force distribution at the cut-plane $z$~=~10~nm (see Fig.~\ref{design}). Magnitude of the lateral force $\mathbf{F}_\|=(F_x,F_y,0)$ is shown with a color scale and the direction with arrows. (b) Distribution of the electric field magnitude in the cut-plane $z$~=~10~nm. Magnitude of the electric field is shown with the color scale. The arrow lines are electric field lines. Panel (c) shows the distribution of the force calculated with a semi-analytical dipolar approach [see Eq.~(\ref{F_analytic})], i.e. where the perturbation of the field by the particle is neglected. Panel (d) shows the numerical simulation of the electric field magnitude distribution without the particle. Panel (a)  should be compared with (c), while (b) with (d). Dark blue shells around nanorods on panels (a), (c) have the width equal to particle's radius. Optical forces are not calculated at those areas, as the nanoparticle's center  cannot approach  the nanorods that close. Upper and lower scales of the color bars correspond to the hyperbolic ($\lambda$~=~600~nm) and elliptic ($\lambda$~=~450~nm) dispersion regimes of the metamaterial, respectively. Electric field amplitude of the incident wave is 1~V/m.}
   \label{main_fig1}
\end{figure*}

The effective permittivity tensor of nanorod metamaterial is given by: 

  \begin{equation}
 \varepsilon=
 \left(
   \begin{matrix} 
      \varepsilon_\bot & 0 & 0\\
      0 & \varepsilon_\bot & 0\\
      0 & 0 & \varepsilon_\| \\
   \end{matrix}
   \right).
  \end{equation}
Here $\varepsilon_\bot$ and $\varepsilon_\|$ are effective permittivities perpendicular and along the wires, respectively.

Dispersion of the tensor components for the structure under consideration was calculated with the approach developed in Ref.~\onlinecite{Elser2006}. The transition between elliptic and hyperbolic dispersion regimes occurs at the wavelength around 523 nm (see Fig.~\ref{fig_eff}). The transition point is called epsilon-near-zero (ENZ) regime, as the real part of the permittivity along the rods is vanishing, if the spatial dispersion effects are ignored. The high density of photonic states as well as the strong scattering emerges in the hyperbolic and ENZ regimes. The wavelength of the external illumination, exploited for optical manipulation in the subsequent investigations, is chosen around this ENZ point in order to distinguish between various dispersion regimes and their impact on optical forces.

\section{Optical force distribution \label{sec:results}}
\subsection{Numerical model \label{sec:num_model}} 

The distribution of the optical forces, acting on the gold nanoparticle placed inside the wire medium is analyzed hereafter. Both the hyperbolic and elliptic dispersion regimes of the bulk metamaterial and their impacts on optical forces are compared and discussed.

The incident wave is chosen to be linearly polarized along the $y$-axis and it propagates along the $z$-axis, as it is shown in Fig.~\ref{design} -- standard normal incidence scenario is considered. Full 3D numerical analysis, based on finite elements method,\cite{pryor2011multiphysics} is performed in order to calculate self-consistent electromagnetic fields in the system. 
Consequently, optical forces acting on the nanoparticle are calculated by integrating the Maxwell's stress tensor components over an imaginary spherical surface surrounding the nanoparticle.

The presence of a single nanoparticle breaks the inherent translation symmetry of the initial metamaterial geometry.  In order to overcome the computation complexity of large systems modeling, Floquet periodical boundary conditions were imposed on finite size geometries. This type of model corresponds to a periodic system with variable unit cell, which consists of a square array of nanorods and the nanoparticle. If the electromagnetic coupling between the particles in adjacent cells is minor, this type of analysis recovers the behavior of the infinite system with a single particle.  

The numerical procedure is as follows: the number of rods in the unit cell is increased gradually and the convergence of a certain quantity (optical forces in our case) is checked. Recently, a similar approach was applied in studies of the Purcell effect in nanorod \cite{slobozhanyuk2015purcell} and wire \cite{mirmoosa2015topological} metamaterials. Square unit cells containing 4, 9, and 16 nanorods were considered and the convergence of optical force values at different points of the metamaterial volume was checked. A unit cell of 4 nanorods (the smallest one) was shown to predict the behavior of an infinite array within the accuracy of several percent. All the subsequent results were obtained for this size of the unit cell. The direct consequence of this calculation is that (i) only nearest neighbor rods define the value of optical force and (ii) nanoparticles in different unit cells almostly do not interact with each other. 

It should be noted, however, that the collective macroscopic behavior of the array is taken into account by imposing periodical Floquet boundary conditions.

\subsection{Lateral force component \label{sec:lateral}}

All the subsequent calculations were done for a particle of 10~nm in diameter. The optical force $\mathbf{F}$, in the most general case, has three non-zero components ($F_x,\ F_y,\ F_z$). The lateral force $\mathbf{F}_\bot=(F_x,F_y,0)$ will be analyzed first. Values of optical forces are normalized to the intensity of the incident wave and volume of the particle in order to perform direct comparisons with other optical manipulation schemes.

The resulting normalized forces at the cut-plane z~=~10~nm, calculated for wavelengths $\lambda$~=~450 and 600~nm, corresponding to the elliptic and hyperbolic dispersion regimes respectively, are shown in Fig.~\ref{main_fig1}(a). Both cases are shown on the same panel, but with different scales of the color bar, as the impact of the excitation wavelength on the spatial distribution of the optical forces is minor. Different color bars indicate the differences in the absolute values. Forces maps at various cut-planes (with different $z$-coordinate) show qualitatively similar behavior. It should be noted, that the values and directions of forces attributed to the geometrical center of the particle, hence certain regions (dark blue shells around nanowires with thickness equal to the radius of nanoparticle) on Fig.~\ref{main_fig1}(a) and (c) are blank, as this center cannot approach the boundaries of the rods.  

It can be seen from Fig.~\ref{main_fig1}(a) that the force distribution has saddle points at the center of the unit cell and at its edges between the rods. These places correspond to the the saddle points of electromagnetic field magnitude distribution [Fig.~\ref{main_fig1}(d)], and, consequently, to the unstable equilibrium positions of the nanoparticle. Some peculiarities in the optical force distribution appear on cut-planes near the edges of the nanorods (z~=~350~nm and z~=~0~nm), but they are attributed to the longitudinal ($z$-component) force component and will be discussed in Sec.~\ref{sec:vertical}.

 \begin{figure*}[t]
   \centering
   \includegraphics[width=0.95\linewidth]{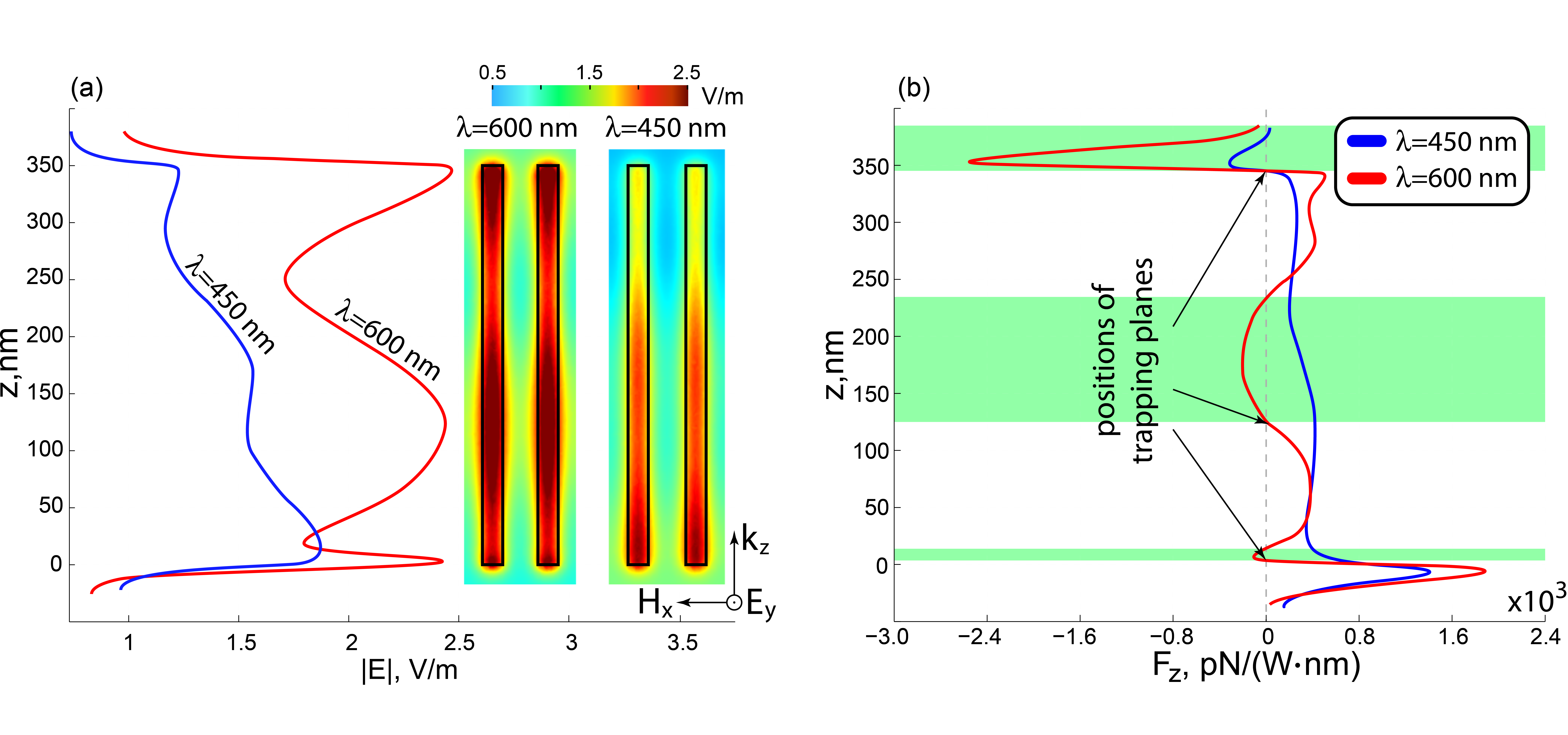} 
   \caption{(a) Distribution of: (a) Total electric field magnitude $|\mathbf{E}|$ and (b) longitudinal optical force component $F_z$ along the line passing through the point with coordinates $x$~=~30~nm and $y$~=~5~nm and parallel to the nanorods for  the elliptic ($\lambda$~=~450~nm) and hyperbolic ($\lambda$~=~600~nm) regimes of the metamaterial. The insets in panel (a) show the distribution of the electric field magnitude in the $zx$-cut-plane passing through $y$~=~5~nm for the elliptic ($\lambda$~=~450~nm) and hyperbolic ($\lambda$~=~600~nm) regimes of the metamaterial. Shaded areas in panel (b) show the region where a pulling force emerges. Electric field amplitude of the incident wave is 1~V/m.}
   \label{fig:Fz}
\end{figure*}

The similarity of the spatial distribution of the forces at hyperbolic and elliptic dispersion  regimes results from the dominating near-field coupling between the nanorods and the particle. Figure~\ref{main_fig1}(b) shows the magnitude distribution of total electric field $|\mathbf{E}|=(E_x^2+E_y^2+E_z^2)^{1/2}$, while the arrows show its direction. One can see that the field map is formed by electrical dipoles induced on the rods and the particle by the incident wave. Orientations of the dipoles nearly coincide with the polarization of the incident wave. Minor deviations from the above description are related to the higher multipole contribution and the interaction between the particle and the nanorods. 

Lateral force distribution analysis can be provided with the following semi-analytical approach. First, the total electric field distribution in the nanorod array under external incident wave without the particle is calculated numerically with the periodic boundary conditions applied. Results of the simulation are shown in Fig.~\ref{main_fig1}(d). The knowledge of the spatial distribution of the electric field magnitude enables calculation of optical forces, with two assumptions: (i) the nanoparticle is represented by a structureless point electric dipole with a moment {\boldmath $\mu$} (ii) the dipole is assumed to act as a small perturbation to the fields of the standalone metamaterial. This means, that  only collective scattering properties of the nanorod array were taken into account, while the mutual re-scattering of the field between the particle and nanorods was neglected. Comparison between Figs.~\ref{main_fig1}(b) and (d) verifies this approximation -- both the structure and values of the field magnitude are similar.

The time averaged optical force acting on the point dipole is given by: \cite{novotny2012principles}

\begin{equation}
\langle {\bf F} \rangle=\frac{\alpha'}{4}\nabla |{\bf E}|^2+\frac{\alpha''}{2}\frac{\omega}{c} \text{Re}\left[{\bf E}\times{\bf H}^*\right],
\label{F_analytic}
\end{equation}
where $\alpha=\alpha'+i\alpha''$  is the complex particle's polarizability. The polarizability of the spherical particle is given by: \cite{novotny2012principles}
\begin{equation}
\alpha=4\pi\varepsilon_0R^3\frac{\varepsilon_{\text{Au}}-1}{\varepsilon_{\text{Au}}+2}.
\end{equation}

The resulting optical force map, calculated using the dipolar approximation [Eq.~(\ref{F_analytic})] appears on Fig.~\ref{main_fig1}(c). The arrows show the direction of the force at the corresponding point. The color pattern corresponds to the absolute value of the force.  The remarkable similarities between Maxwell's stress tensor calculations  [Fig.~\ref{main_fig1}(a)] and the approximate analytical model [Fig.~\ref{main_fig1}(c)] suggest the validity of the dipolar model and highlights the impact of near-fields on the optical force. It should be noted, however, that overall values of optical forces, calculated within those approaches, have about 20\% difference, which is related to the finite size of the particle and the mutual field re-scattering between the particle and nanorods.

 \begin{figure*}[t]
   \centering
   \includegraphics[width=0.65\linewidth]{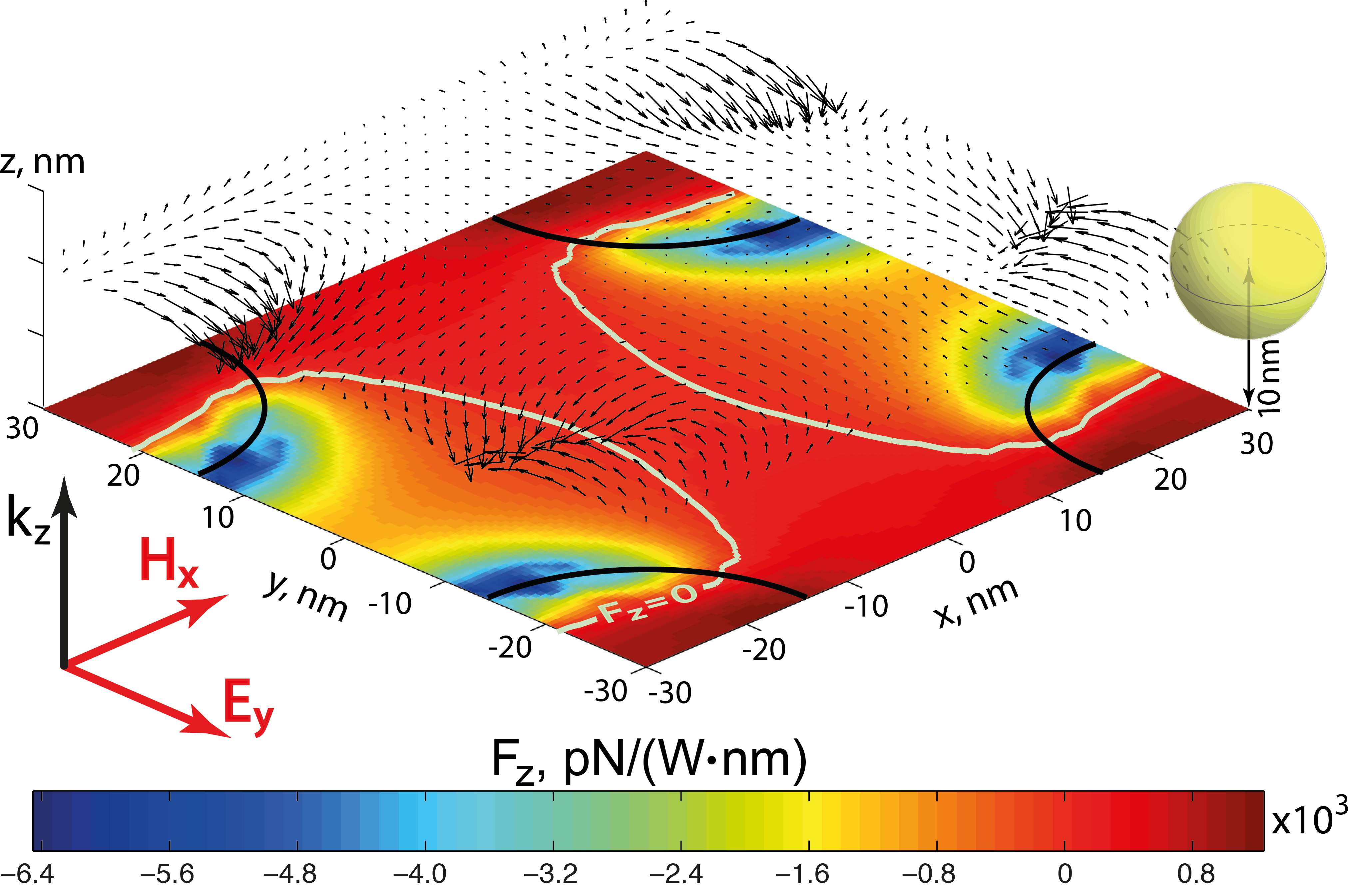} 
   \caption{(color online) Distribution of the normalized optical force acting on the nanoparticle, situated at the lateral plane above the metamaterial. Distance between the center of the nanoparticle and top faces of the nanorods is 10~nm. Arrows show the direction of the optical force. Color map shows the distribution of the optical force component parallel to the nanorods ($F_z$). Black solid lines show the the geometrical edges of the nanorods.}
   \label{fig_attraction_force}
\end{figure*}

\subsection{Vertical force component \label{sec:vertical}} 

The distribution of the lateral optical force component (perpendicular to the nanorods) was analyzed in the previous section. Longitudinal force component $F_z$ (parallel to the nanorods) is analyzed here.

As it was already mentioned, the homogenization procedure averages the near-fields over the unit cell, hence, it is inapplicable for estimation of gradient optical force in the lateral plane. Nevertheless, the field distribution along the $z$-axis can be roughly estimated considering the slab of the nanorod metamaterial as a Fabry-Perot resonator in $z$-direction.\cite{slobozhanyuk2015purcell} Therefore, it is reasonable to expect a standing wave in the slab (along the $z$-axis) and maxima of the electric field resulting in in-plain trapping of the particle. The results of the numerical calculation suggest the validity of this hypothesis. Estimation of the field maxima position deeply inside the slab can be provided by the effective medium approximation \cite{Silveirinha2006,tyshetskiy2014guided} but a more detailed analysis, that takes into account boundary effects, demands numerical simulation.

The profile of the total electric field magnitude along the line parallel to the rods and passing through the point $x$~=~30~nm and $y$~=~5~nm calculated without nanoparticle for the elliptic ($\lambda$~=~450~nm) and hyperbolic ($\lambda$~=~600~nm) regimes is shown in Fig.~\ref{fig:Fz}(a). The insets show distribution of the total electric field magnitude in $xz$-plane passing through $y$~=~5~nm. One can see that electric field distribution along the $z$-axis strongly depends on the wavelength of the incident wave. In the hyperbolic regime ($\lambda$~=~600~nm), three distinct field maxima are observed -- one inside the slab and two in the vicinity of its boundaries. In the elliptic regime ($\lambda$~=~450~nm), field decays inside the metamaterial and weak oscillations do not possess sharp field maxima. Additional contribution to those differences (apart from the interplay of dispersion regime and geometry, namely Fabry-Perot conditions) comes from a strong wavelength dependence of losses in gold:\cite{Johnson1972}

\begin{equation}
\left|\frac{\text{Re}(\varepsilon_\text{Au})}{\text{Im}(\varepsilon_\text{Au})}\right|\approx
\begin{cases}
      0.3  \ \ \ \text{for} \ \lambda~=~450~\text{nm}; \\
      6.2  \ \ \ \text{for} \ \lambda~=~600~\text{nm}. 
\end{cases}
\label{Eq_FOM}
\end{equation} 
Optical losses cause the reduction in quality factors of the modes, smearing out the sharp peaks, as could be seen in the case of elliptic dispersion.

Distributions of the $z$-component of the optical force along the nanorod for elliptic ($\lambda~=~450$~nm) and hyperbolic ($\lambda~=~600$~nm) regimes are shown in Fig.~\ref{fig:Fz}(b). Positions of the stable trapping in transverse planes are marked with arrows (note, that the force derivative should be negative in order to obtain a stable equilibrium). Shaded areas on the figure show the regions within the metamaterial, where the optical force component $F_z$ is directed towards the light source (for $\lambda~=~600$~nm). Optical pulling forces or optical attraction gained considerable attention over the last decade, as it provides additional flexible degree of freedom in optical manipulation.\cite{chen2011optical}

The hyperbolic regime supports three regions of optical attraction, while the elliptic has only one, as could be seen in Fig.~\ref{fig:Fz}(b). This occurrence could be understood as follows: in the elliptic regime both weak gradient of the electric field magnitude [see Fig.~\ref{fig:Fz}(a)] and high material losses of the particle result in the domination of radiation pressure [second term in Eq.~(\ref{F_analytic})] over the gradient force. The radiation pressure is co-directional with the Poynting vector of the incident radiation, so the optical attraction cannot be obtained (unless negative-index materials are involved, which is not the case here). Nevertheless, the first term of Eq.~(\ref{F_analytic}) overcomes the second one in the vicinity of the nanorod's edge where strong gradient of the electric field magnitude is observed [see Fig.\ref{fig:Fz}(a)]. In the hyperbolic regime, on the other hand, there are several regions where the optical force component is directed to the light source -- that's the result of high quality factor Fabry-Perot modes and dominating real part of the particle's polarizability.



As a separate case, the particle situated over the metamaterial slab will be considered next. This scenario describes the case where the metamaterial is used as a substrate for advanced optical manipulation. Results of numerical studies appear in Fig.~\ref{fig_attraction_force}, showing the distribution of the vertical optical force acting on the nanoparticle in the lateral plane of $z$~=~10~nm above the nanorods.  It could be seen, that the maximal attraction force on the particle emerges in the vicinity of nanorods edges (the sample is illuminated from below -- see Fig.~\ref{fig_attraction_force}). Arrows indicate the direction of the optical force. Color pattern shows the distribution of the optical force component parallel to the nanorods ($F_z$).  Solid white lines correspond to $F_z$~=~0. Remarkable behaviour of forces above the metamaterial substrate could suggest the later as an auxiliary nanostructure or metasurface, providing additional flexibility on optical manipulation.
 

\section{Conclusion \label{sec:conclusion}}

In this work, comprehensive analysis of the optical forces acting on a metal nanoparticle placed inside or in the vicinity of three-dimensional nanorod metamaterial slab was performed. Numerical simulations of finite size square unit cells with periodical Floquet boundary conditions enable to take into account all collective effects in the metamaterial and estimate optical forces on small particles. Unit cells containing 4, 9, and 16 nanorods were analyzed and the convergence of the optical forces for different positions of the particle was checked. It was shown that the smallest unit cell already reproduces the effect of optical forces on a particle, situated within the infinite metamaterial. Therefore, only four neighboring nanorods nearest to the particle make the dominant contribution to the optical forces. This statement has been confirmed with the developed semi-analytical model which neglects the particle's interior and the re-scattering effects between the particle and nanorods. Furthermore, it was shown that the 'topological transition' from the elliptic to hyperbolic dispersion regime of the metamaterial, usually having an impact on various light-matter interaction processes, is less important for optical forces. 

In-plane optical trapping and optical pulling forces were observed. The comprehensive numerical modeling enables estimation of optical forces values, normalized to incident power and particle's volume. Values as high as 2.3$\times$10$^3$~pN/W/nm for both lateral and optical pulling forces were predicted. Those results overcome routinely reported values (see, e.g., Ref.~\onlinecite{yang2011optical}) by an order of magnitude and are comparable with other advanced layouts (see, e.g., Ref.~\onlinecite{Wang2009}).

The remarkable structure of predicted optomechanical interactions (in particular pulling forces), mediated by the metamaterial, make the later to be a promising platform for large span on multidisciplinary applications, involving demands for precise nanoscale mechanical manipulation, including trapping sorting, mixing and more.
\\ 
\acknowledgements
This work was partially supported by the Government of the Russian Federation (Grant 074-U01), by the Russian Foundation for Basic Research (No. 15-02-01344), by the Program on Fundamental Research in Nanotechnology and Nanomaterials of the Presidium of the Russian Academy of Sciences, and by the Russian Science Foundation (Grant No. 14-12-01227). AAB thanks Russian Federation President support program of leading scientific schools (NSh-5062.2014.2) and RFBR (No. 14-02-01223).

\bibliography{references}
\end{document}